\documentclass[fleqn,usenatbib]{mnras}

\usepackage{newtxtext,newtxmath}
\usepackage{makecell}

\usepackage[T1]{fontenc}

\DeclareRobustCommand{\VAN}[3]{#2}
\let\VANthebibliography\thebibliography
\def\thebibliography{\DeclareRobustCommand{\VAN}[3]{##3}\VANthebibliography}

\usepackage{graphicx}	
\usepackage{amsmath}	

\newcommand{\here}{\href{https://github.com/gannonjs/Published_Data/tree/main/UDG_Spectroscopic_Data}{here}}
\newcommand{\website}{\href{https://github.com/gannonjs/Published_Data/tree/main/UDG_Spectroscopic_Data}{https://github.com/gannonjs/Published\_Data/tree/main/UDG\_Spectroscopic\_Data}}

\title[Spectroscopic UDG Catalogue]{A Catalogue and Analysis of Ultra-Diffuse Galaxy Spectroscopic Properties}

\author[J. S. Gannon et al.]{
Jonah S. Gannon,$^{1, 2}$\thanks{E-mail: jonah.gannon@gmail.com}
Anna Ferr\'e-Mateu,$^{3,4,1}$
Duncan A. Forbes,$^{1, 2}$
Jean P. Brodie,$^{1, 2, 5}$
Maria Luisa Buzzo,$^{1, 2}$
\newauthor and Aaron J. Romanowsky$^{6, 5}$
\\
$^{1}$ Centre for Astrophysics and Supercomputing, Swinburne University, John Street, Hawthorn VIC 3122, Australia.\\
$^{2}$ ARC Centre of Excellence for All Sky Astrophysics in 3 Dimensions (ASTRO 3D)\\
$^{3}$ Instituto de Astrof\'isica de Canarias, Calle V\'ia L\'actea S/N, E-38205, La Laguna, Tenerife, Spain\\
$^{4}$ Departamento de Astrofísica, Universidad de La Laguna, 38206, La Laguna (S.C. Tenerife), Spain\\
$^{5}$ Department of Astronomy \& Astrophysics, University of California Santa Cruz, 1156 High Street, Santa Cruz, CA 95064, USA\\
$^{6}$ Department of Physics and Astronomy, San Jos\'e State University, One Washington Square, San Jose, CA 95192, USA
}

\date{Accepted XXX. Received YYY; in original form ZZZ}

\pubyear{2024}

\begin{document}
\label{firstpage}
\pagerange{\pageref{firstpage}--\pageref{lastpage}}
\maketitle

\begin{abstract}
In order to help facilitate the future study of ultra-diffuse galaxies (UDGs) we compile a catalogue of their spectroscopic properties. Using it, we investigate some of the biases inherent in the current UDG sample that have been targeted for spectroscopy. In comparison to a larger sample of UDGs studied via their spectral energy distributions (SED), current spectroscopic targets are intrinsically brighter, have higher stellar mass, are larger, more globular cluster-rich, older, and have a wider spread in their metallicities. In particular, many spectroscopically studied UDGs have a significant fraction of their stellar mass contained within their globular cluster (GC) system. We also search for correlations between parameters in the catalogue. Of note is a correlation between alpha element abundance and metallicity as may be expected for a `failed galaxy' scenario. However, the expected correlations of metallicity with age are not found and it is unclear if this is evidence against a `failed galaxy' scenario or simply due to the low number statistics and the presence of outliers. Finally, we attempt to segment our catalogue into different classes using a machine learning K-means method. We find that the clustering is very weak and that it is currently not warranted to split the catalogue into multiple, distinct sub-populations. Our catalogue is available online and we aim to maintain it beyond the publication of this work.
\end{abstract}


%



\section{Introduction}
While low surface brightness (LSB) galaxies have been studied for decades now \citep{Reaves1962, Disney1976, Sandage1984, Bothun1987, Impey1988, Impey1997, Dalcanton1997, Conselice2003} recent discoveries suggest that many more LSB galaxies exist than was first expected. In particular, the work of \citet{vanDokkum2015} has raised interest in so-called `ultra-diffuse galaxies' (UDGs) after they reported forty-seven such examples in the Coma Cluster. They defined these galaxies to be those with surface brightness, $\mu_{g,0}>24$~mag arcsec$^{-2}$ and half-light radii, $R_{\rm e}>1.5$ kpc. Thousands more examples of UDGs have been found across all environments (e.g., \citealp{Yagi2016, MartinezDelgado2016, vanderBurg2017, Roman2017, Roman2017b, Roman2019, Janssens2017, Janssens2019, Muller2018, Prole2019, Forbes2019, Forbes2020b, Zaritsky2019, Zaritsky2021, Barbosa2020}). It now appears that $>7$\% of all galaxies may be ultra-diffuse \citep{Li2023}. Elucidating UDG formation is thus a key research topic for those wishing to understand galaxy formation. 

A multitude of theories exist to explain UDG formation. These mostly rely on either external (e.g., tidal heating, tidal stripping, environmental quenching, ram pressure stripping, galaxy mergers; \citealp{Carleton2019, Sales2020, Doppel2021, Wright2021, Jones2021, vanDokkum2022}) or internal (e.g., high dark matter halo spin, stellar feedback, stellar passive evolution; \citealp{DiCintio2017, Amorisco2016, Rong2017, Benavides2023, Chan2018, Fielder2024}) processes. Combinations of both are also possible (e.g., \citealp{Jiang2019, Martin2019, Sales2020}). 

Crucially the different proposed formation mechanisms are expected to leave different imprints in the stellar populations and dark matter halo properties of the resulting UDG. To provide a pair of contrasting examples: 1) a UDG forming via episodic stellar feedback is expected to have an extended star formation history, a dwarf-like metallicity and a normal, dwarf-like dark matter halo (and thus lower velocity dispersion and globular cluster counts); while 2) a UDG forming at high redshift and quenching quickly is expected to have an old stellar population reflective of a single burst of star formation at high redshift, low metallicities reflective of the lack of time for chemical enrichment in the stellar population and a more massive dark matter halo (and thus higher velocity dispersion and globular cluster counts). Galaxies with properties resembling dwarf galaxies have been dubbed `puffy dwarfs' in the literature due to their resemblance to the large-end tail of the dwarf half-light radius -- luminosity relation (e.g., the UDGs forming via strong stellar feedback discussed above). Galaxies that have properties resembling a formation at high redshift and catastrophic quenching have been dubbed `failed galaxies' in the literature \citep{vanDokkum2015, Danieli2022, Forbes2024}. Differentiating between these properties, and thus the corresponding formation scenario, may be accomplished through spectroscopy of the UDG's stellar body (e.g., \citealp{FerreMateu2023}). 

Alternatively, some of the properties (e.g., age/metallicity/star formation timescales) desired for elucidating UDG formation scenarios may be measured using spectral energy distribution (SED) fitting (e.g., \citealp{Barbosa2020, Buzzo2022}). This has the advantage of allowing larger samples of UDGs to be studied. Recent results from the SED fitting of UDGs have been able to separate them into two distinct classes using a K-means clustering analysis \citep{Buzzo2024}. Interestingly the mean properties of these classes were found to agree with the `puffy dwarf'/`failed galaxy' examples given above. To date, no similar analysis has been performed on spectroscopic UDG samples. 

Spectroscopy is extremely time intensive, requiring multiple hours on the world's largest optical telescopes ($\geq8$m-class). As such, spectroscopic studies of UDG velocity dispersions and stellar populations tend to be limited to single objects and/or small samples (e.g., \citealp{vanDokkum2017, Toloba2018, Gu2018, Alabi2018, FerreMateu2018, RuizLara2018, MartinNavarro2019, Emsellem2019, Danieli2019, vanDokkum2019b, Chilingarian2019, Muller2020, Gannon2020, Gannon2021, Gannon2022, Gannon2023, Forbes2021}). This has led to a UDG literature that requires significant effort to compile whenever a new object is studied and comparisons are wanted to previously published works. It has also led to a lack of understanding as to the selection biases of the current spectroscopic sample, which many UDG formation conclusions are based on.

In this work, we provide a compilation of current UDG spectroscopic properties in a single catalogue for easy access. In Section \ref{sec:inclusion} we present the criteria for galaxies that have been included in our catalogue. In Section \ref{sec:catalogue} we present the catalogue with individual galaxy notes. In Section \ref{sec:discussion} we provide a brief discussion of our sample in comparison to a large sample of UDGs \citet{Buzzo2024}, investigate correlations in the sample and study its GC-richness. In Section \ref{sec:housekeeping} we provide some housekeeping details including referencing preferences and catalogue availability. We intend to keep the catalogue updated beyond the publication of this paper. Finally, a brief summary and conclusions are presented in Section \ref{sec:conclusions}.

\section{Inclusion Criterion} \label{sec:inclusion}
In order to be included in this catalogue we require the galaxy to be both 1) a UDG and 2) have spectroscopically measured mass-weighted stellar ages and metallicities and/or a spectroscopically measured stellar/globular cluster (GC) velocity dispersion.

For the UDG definition, we wished to follow the original UDG definition ($\mu_{g,0}>24$~mag arcsec$^{-2}$ and $R_{\rm e}>1.5$ kpc; \citealp{vanDokkum2015}) but derive it in the $V$-band, to make it easier to search for UDGs in established catalogues such as those of \citet{mcconnachie2012} for the Local Group. We also convert from a central surface brightness ($\mu_{g,0}$) to an average surface brightness within the half-light radius ($\langle \mu_{V} \rangle_{\rm e}$). We therefore take the original definition and apply the colour correction $V = g - 0.3$ along with an aperture correction of $\langle \mu \rangle_{\rm e} = \mu_{0} + 1$. Our aperture correction is based on equations 7 and 9 in \citet{Graham2005} for a galaxy of S\'ersic index ($n$) slightly below 1, which is representative of a large population of UDGs in e.g., the Coma Cluster \citep{Yagi2016}. We, therefore, derive our UDG surface brightness criterion as:

\begin{equation}
    \langle \mu_{V} \rangle_{\rm e} = \mu_{g,0} + 1 - 0.3 = 24.7 \mathrm{mag~arcsec^{-2}}
\end{equation}

We make no changes to the half-light radius criterion from the original \citet{vanDokkum2015} definition, keeping a semi-major half-light radius $R_{\rm e}>1.5$~kpc. 

To be specific our final galaxy inclusion criteria for this catalogue are:

\begin{enumerate}
    \item An average $V$-band surface brightness within the half-light radius of $\langle \mu_{V} \rangle_{\rm e} > 24.7$~mag arcsec$^{-2}$.
    \item A semi-major half-light radius $R_{\rm e}>1.5$~kpc.
    \item Either a spectroscopically measured velocity dispersion and/or a mass-weighted stellar age and metallicity 
\end{enumerate}

It is worth noting that different UDG definitions can bias the inferred different formation pathways \citep{vanNest2022} and the UDG definition itself may bias the sample to redder galaxies than one that searches for large-size outliers \citep{Li2023}. In addition, our choice of a mean surface brightness within the half-light radius may include a small percentage of higher-S\'ersic index galaxies that a central surface brightness definition would exclude (see e.g., \citealp{Greco2018b} fig. 6). 

\section{Catalogue and Individual Galaxy Notes} \label{sec:catalogue}
We present the full catalogue in Appendix A, Table \ref{tab:cropped_table1}, Table \ref{tab:cropped_table2} and Table \ref{tab:cropped_table3} as well as online \here\footnote{\website}. When the mean $V-$band surface brightness within the half-light radius was unavailable it was calculated using the magnitude, half-light radius and equation 11 of \citet{Graham2005}.  When magnitudes/surface brightnesses were only available in $g-$band the magnitude has been transformed from $g$-band using $V=g-0.3$. Unless otherwise stated, when multiple measurements were available for the same property they were combined with weighting according to their uncertainties. Below we list individual notes for each UDG we have included in the catalogue.

\begin{figure*}
    \centering
    \includegraphics[width = 0.98 \textwidth]{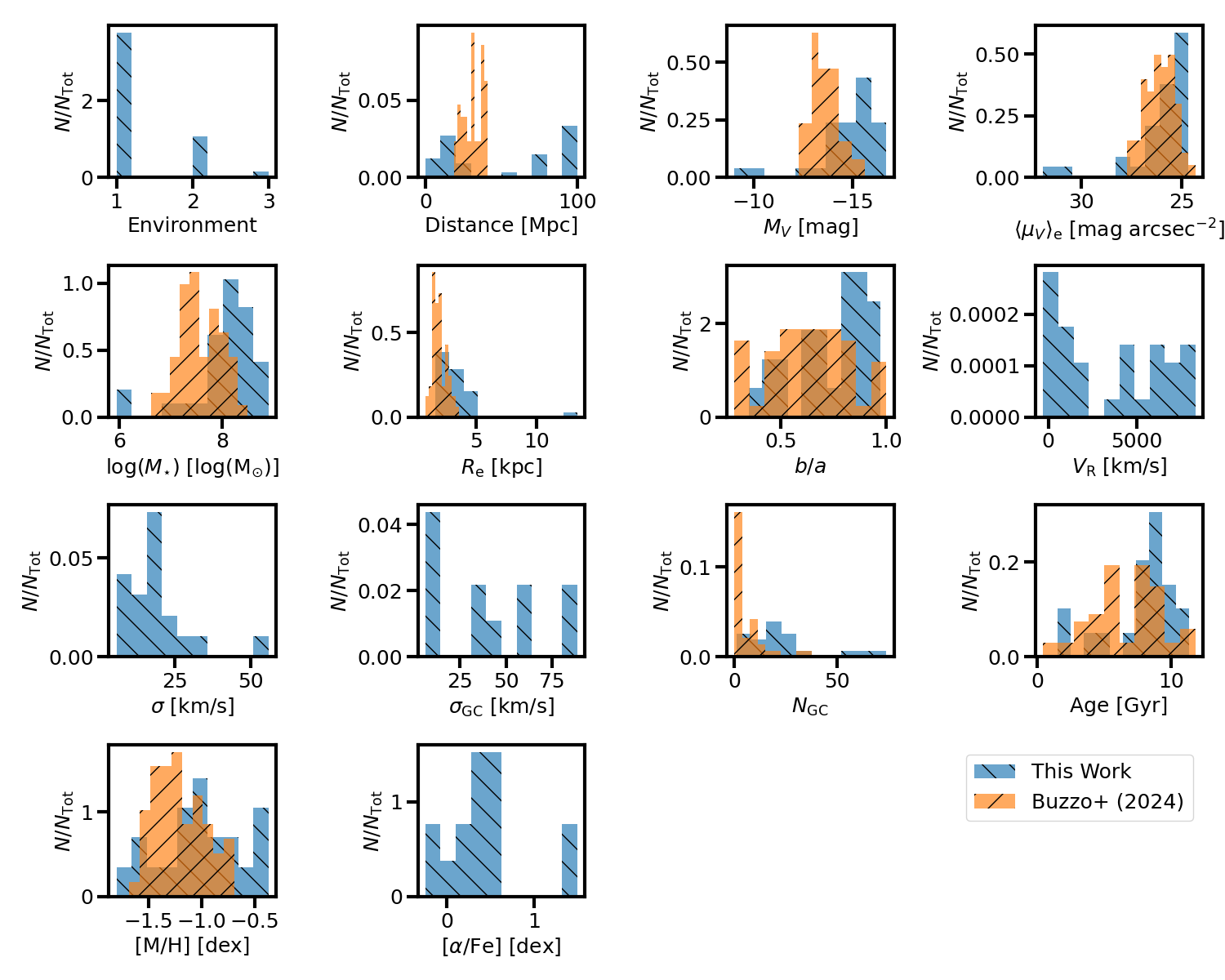}
    \caption{Histograms of each of the UDG properties in the catalogue. From left to right, top to bottom these are: 1) Environment, where 1=cluster, 2=group and 3=field, 2) Distance to the UDG, 3) The $V$-band absolute magnitude, 4) The average $V$-band surface brightness within the half-light radius, 5) Total stellar mass, 6) 2D projected, semi-major half-light radius, 7) Axial ratio $b/a$, 8) Recessional velocity, 9) Stellar velocity dispersion, 10) GC system velocity dispersion, 11) Number of GCs, 12) Mass-weighted stellar age, 13) Mass-weighted stellar metallicity and 14) Stellar alpha abundance ([$\alpha$/Fe]). The catalogue data are plotted in blue. In orange we include results from the SED fitting of MATLAS Survey UDGs from the study of \citet{Buzzo2024}. It is worth noting that for all of the SED sample, and the majority of the spectroscopic catalogue, the distance is assumed based on the environmental association. This assumption will affect several other panels that are dependent on the distance to derive physical units. In comparison to the larger SED sample, current spectroscopically studied UDGs tend to be intrinsically brighter, have higher stellar masses, are larger, more GC-rich, older and to have a wider spread in their metallicities. }
    \label{fig:histograms}
\end{figure*}

\subsection{Andromeda XIX}
Andromeda XIX is a satellite of M31 and resides in the Local Group. Due to its extremely low surface brightness, it is unlikely similar analogues may be found outside of the Local Group. We note that Andromeda XIX is likely affected by tidal processes interacting with the nearby M31 \citep{Collins2020, Collins2022}. Any dynamical masses calculated with the data in the catalogue should be interpreted with caution. Due to the extremely diffuse nature of this object, the half-light radius, magnitude and surface brightness are highly uncertain. The listed stellar mass was calculated from the $V$-band magnitude in \citet{Martin2016} assuming $M_{\star}/L_{V}$ = 2. The data for this galaxy are taken from the works of \citet{Martin2016}, \citet{Collins2020} and \citet{Gannon2021}.

\subsection{Antlia II}
Antlia II is a satellite of the Milky Way and resides in the Local Group. Due to its extremely low surface brightness, it is unlikely that similar analogues will be found outside of the local group. Dynamical modelling by \citet{torrealba2019} suggests that a combination of tidal stripping and a cored dark matter profile can explain the properties of Antlia II. Due to the suggestion of tidal stripping, any dynamical mass calculated with the data should be treated with caution. The data for this galaxy are taken from the works of \citet{mcconnachie2012} and \citet{torrealba2019}.

\subsection{DF44}
DF44 is in the Coma cluster and has been one of the best-studied UDGs to date. It is one of only two UDGs that has had spatially resolved kinematic and stellar population gradients measured (the other being NGC~1052-DF2). This interest has mostly been the result of claims of a rich GC system associated with the galaxy \citet{vanDokkum2017} although there is currently some disagreement on the total GC numbers of DF44 in the literature \citep{Saifollahi2021, Saifollahi2022}. See \citet{Forbes2024} for a further discussion of these numbers. Following this work, we choose the \citet{vanDokkum2017} GC number. When quoting the $N_{\rm GC}$ from \citet{vanDokkum2017} we use the number listed in their abstract (74$\pm$18) which is slightly different to that in Table 1. We have been advised this is the correct number (P. van Dokkum, private communication). While we classify DF44 as being in the Coma cluster, its phase space positioning suggests it may just be beginning to infall as part of a small group \citep{vanDokkum2019b}. As such, some authors have classified it with low-density UDGs when considering its formation (e.g., \citealp{FerreMateu2023}).  The radial velocity was derived using $V_{\rm r} = c \times \ln{(1+z)}$ from the redshift listed in footnote 6 of \citet[$z=$0.02132]{vanDokkum2017}. The data for this galaxy are taken from the works of \citet{vanDokkum2016, vanDokkum2017, vanDokkum2019b, Gannon2021, Villaume2022, Webb2022} and \citet{Saifollahi2022}.

\subsection{DF07}
DF07 is in the Coma Cluster. The GC count is a combination of values by \citet[39.1$\pm$23.8]{Lim2018} and \citet[22$^{+5}_{-7}$]{Saifollahi2022}. The data for this galaxy are taken from the works of \citet{vanDokkum2015, Gu2018, Lim2018, Saifollahi2022} and \citet{FerreMateu2023}.

\subsection{DF17}
DF17 is in the Coma Cluster. The GC count is a combination of values by \citet[28$\pm$14]{Peng2016}, \citet[27$\pm$5]{Beasley2016b}, \citet[25$\pm$11]{vanDokkum2017} and \citet[26$^{+17}_{-7}$]{Saifollahi2022}. All values are within uncertainties of one another and are in good agreement \citep{Forbes2024}. The data for this galaxy are taken from the works of \citet{Peng2016, Beasley2016b, vanDokkum2017, Gu2018} and \citet{Saifollahi2022}.

\subsection{DF26}
DF26 is a Coma cluster galaxy. This galaxy is also known as Y093 or Yagi 093. The magnitude was calculated from $R$-band using $V = R + 0.5$ (based on Virgo dEs and Coma LSBs; \citealp{vanZee2004, Alabi2020}). Light-weighted ages and metallicities are available for this galaxy from \citet{RuizLara2018}. The data for this galaxy are taken from the works of \citet{Yagi2016, Alabi2018, Lim2018} and \citet{FerreMateu2018}. 

\subsection{DFX1}
DFX1 is in the Coma Cluster. There is currently some disagreement on the total GC numbers of DF X1 in the literature \citep{Saifollahi2021, Saifollahi2022}. See further \citet{Forbes2024} for a discussion of these numbers. Following this work, we choose the \citet{vanDokkum2017} GC number. When quoting the $N_{\rm GC}$ from \citet{vanDokkum2017} we use the number listed in their abstract which is slightly different from the number in Table 1. The radial velocity was derived using $V_{\rm R} = c \times \ln{(1+z)}$ from the redshift listed in section 2.1 of \cite{vanDokkum2017}. Note that it is likely that the stellar velocity dispersion is also affected by the barycentric correction issue described in footnote 16 of \cite{vanDokkum2019b}, however, the effect is likely small (P. van Dokkum, private communication). The data for this galaxy are taken from the works of \citet{vanDokkum2017}, \citet{Gannon2021}, \citet{Saifollahi2022} and \citet{FerreMateu2023}. 

\subsection{DGSAT-$\mathrm{I}$}
DGSAT-$\mathrm{I}$ is listed as field although we note that it is located near the Pisces--Perseus supercluster and may potentially be a `backsplash' galaxy \citep{MartinezDelgado2016, Papastergis2017, Benavides2021}. The backsplash galaxy hypothesis has been disfavoured by \citet{Janssens2022} and thus we continue to list this galaxy as a field object. Note that some of the GCs counted are more luminous than expected given a traditional GC luminosity function \citep{Janssens2022}. The data for this galaxy are taken from the works of \citet{MartinezDelgado2016, MartinNavarro2019} and \citet{Janssens2022}. 

\subsection{Hydra $\mathrm{I}$ UDG 11}
Hydra $\mathrm{I}$ UDG 11 is in the Hydra $\mathrm{I}$ cluster. The magnitude was converted to $g$ band using the listed $g-r$ colour in \citet{Iodice2020} and then transformed to $V$-band assuming $V = g-0.3$. The data for this galaxy are taken from the works of \citet{Iodice2020} and \citet{Iodice2023}.

\subsection{J130026.26+272735.2}
This UDG is in the Coma Cluster. The magnitude and surface brightness were calculated from $R$-band using $V=R+0.5$ (based on Virgo dEs and Coma LSBs; \citealp{vanZee2004, Alabi2020}). The data for this galaxy are taken from the work of \citet{Chilingarian2019}.

\subsection{NGC 1052-DF2}
We classify NGC 1052-DF2 as being in the NGC~1052 group. However, there is the possibility that it is no longer bound to the NGC~1052 group as a result of its formation mechanism (e.g., \citealp{Shen2021, vanDokkum2022}). NGC 1052-DF2 is irregular for a galaxy in having both an extremely low measured velocity dispersion \citep{vanDokkum2018, Danieli2019} and an excess of bright GCs beyond what is expected given the established GC luminosity function for normal galaxies \citep{vanDokkum2018b, Shen2021}. The addition of a weak rotational component, as allowed by the data, may help alleviate the paucity of dark matter suggested by its velocity dispersion \citep{Emsellem2019, Lewis2020, Montes2021}. Furthermore, it may currently be undergoing a tidal interaction (\citealp{Keim2021}, although see \citealp{Montes2021, Golini2024}). We note that there existed some initial controversy over the distance to NGC~1052-DF2, whereby a smaller distance can solve much of the galaxy's irregular properties (see e.g., \citealp{Trujillo2019, Monelli2019}). This controversy is now largely resolved by the deep \textit{HST} imaging of \citet{Shen2021}, with this distance being further updated in Appendix A of \citet{Shen2023}.

We adopt the recessional velocity and velocity dispersion measurements reported from the Keck/KCWI data of \citet{Danieli2019} over those reported from the VLT/MUSE data of \citet{Emsellem2019} due to Keck/KCWI having the higher instrumental resolution. When quoting GC counts, we use the number of GCs measured by \citet{Shen2021} in the traditional GC luminosity function luminosity range, which excludes the brighter GC sub-population. We adopt the stellar population properties reported from VLT/MUSE data in \citet{Fensch2019} over those reported from GTC/OSIRIS data in \citet{RuizLara2019} due to the larger field of view of VLT/MUSE being able to measure a more global value for the galaxy. Both values are in agreement. The data for this galaxy are taken from the works of \citet{vanDokkum2018, Fensch2019, Danieli2019, Shen2021} and \citet{Shen2023}.

\subsection{NGC 5846$\_$UDG1}
NGC 5846$\_$UDG1 is in the NGC~5846 group. This galaxy is also known as MATLAS-2019 \citep{Muller2020} and as NGC~5846-156 by \citet{Mahdavi2005}. Here, we have adopted the velocity dispersion and redshift from \cite{Forbes2021} rather than those measured in \cite{Muller2020} due to the higher instrumental resolution in the data used by \citet{Forbes2021}. We additionally adopt the distance/GC richness from \cite{Danieli2022} rather than that reported in \cite{Muller2021} due to the greater depth of the \textit{HST} data. The data for this galaxy are taken from the works of \citet{Forbes2019, Muller2020, Muller2021, Forbes2021, Danieli2022} and \citet{FerreMateu2023}.

\subsection{NGVSUDG-19}
NGVSUDG-19 is in the Virgo cluster. The data for this galaxy are taken from the works of \citet{Lim2020} and \citet{Toloba2023}.

\subsection{NGVSUDG-20}
NGVSUDG-20 is in the Virgo cluster. The data for this galaxy are taken from the works of \citet{Lim2020} and \citet{Toloba2023}.

\subsection{PUDG-R15}
PUDG-R15 is in the Perseus cluster. The data for this galaxy are taken from the works of \citet{Gannon2022} and \citet{FerreMateu2023}. 

\subsection{PUDG-R16}
PUDG-R16 is in the Perseus cluster. The data for this galaxy are taken from the work of \citet{Gannon2022}.

\subsection{PUDG-R84}
PUDG-R84 is in the Perseus cluster. The data for this galaxy are taken from the works of \citet{Gannon2022} and \citet{FerreMateu2023}. 

\subsection{PUDG-S74}
PUDG-S74 is in the Perseus cluster. The data for this galaxy are taken from the works of \citet{Gannon2022} and \citet{FerreMateu2023}.

\subsection{Sagittarius dSph}
The Sagittarius dSph is a satellite of the Milky Way in the Local Group and is known to be completely tidally disrupted around the Milky Way \citep{Ibata2001}. Any mass calculated with values listed in the catalogue should be treated with extreme caution due to the lack of equilibrium in the galaxy. The data for this galaxy are taken from the works of \citet{mcconnachie2012, Karachentsev2017} and \citet{Forbes2018}.

\subsection{UDG1137+16}
UDG1137+16 is a satellite of the galaxy UGC~6594 in a group environment. It is also known as dw1137+16 by \citet{Muller2018}. It has a disturbed morphology suggestive that it is undergoing stripping \citep{Gannon2021}. Any mass calculated with the values listed in the catalogue should be treated cautiously. $M_r$ was transformed into $V$-band using stated $g-r$ colour (0.65) and $V=g-0.3$. The data for this galaxy are taken from \citet{Gannon2021} and \citet{FerreMateu2023}.

\subsection{VCC 1017}
VCC 1017 is a Virgo cluster galaxy. The data for this galaxy are taken from the works of \citet{Lim2020} and \citet{Toloba2023}.

\subsection{VCC 1052}
VCC 1052 is a Virgo cluster galaxy. It has been noted to have a peculiar morphology with the possibility of spiral arms and/or tidal features \citep{Lim2020}. The data for this galaxy are taken from the works of \citet{Lim2020} and \citet{Toloba2023}.

\subsection{VCC 1287}
VCC 1287 is a Virgo cluster galaxy. Here the GC velocity dispersion is a combination of that measured by \citet[33$^{+16}_{-10}$]{Beasley2016} and \citet[39$^{+20}_{-12}$]{Toloba2023}. Both values agree within uncertainties. The data for this galaxy are taken from the works of \citet{Beasley2016, Gannon2020, Gannon2021, Lim2020} and \citet{Toloba2023}.

\subsection{VCC 615}
VCC 615 is a Virgo cluster galaxy. The data for this galaxy are taken from the works of \citet{Lim2020} and \citet{Toloba2023}.

\subsection{VCC 811}
VCC 811 is a Virgo cluster galaxy. The data for this galaxy are taken from the works of \citet{Lim2020} and \citet{Toloba2023}.

\subsection{VLSB-B}
VLSB-B is a Virgo cluster galaxy. Note that many of the properties presented in the catalogue were updated in \cite{Toloba2023} from those listed in \cite{Toloba2018}. The data for this galaxy are taken from the works of \citet{Toloba2018, Lim2020} and \citet{Toloba2023}. 

\subsection{VLSB-D}
VLSB-D is a Virgo cluster galaxy. It has an elongated structure and velocity gradient \citep{Toloba2018} that suggests it is undergoing tidal stripping. Any dynamical mass derived with the properties listed must be treated with caution. Note that many of the properties presented in the catalogue were updated in \cite{Toloba2023} from those listed in \cite{Toloba2018}. It is worth noting that while this galaxy has an estimated GC number of 13 $\pm$~6.9, 14 GCs have been confirmed spectroscopically. The data for this galaxy are taken from the works of \citet{Toloba2018, Lim2020} and \citet{Toloba2023}. 

\subsection{WLM}
WLM is a galaxy on the outskirts of the Local Group. It is gas-rich and undergoing active star formation \citep{Leaman2009}. It also likely has a large rotation component in its dynamics \citep{Leaman2009}.
The data for this galaxy are taken from \citet{mcconnachie2012} and \citet{Forbes2018}. 

\subsection{Yagi 098}
Yagi 098 is a Coma cluster galaxy. The magnitude was calculated from $R$-band using $V = R + 0.5$ (based on Virgo dEs and Coma LSBs; \citealp{vanZee2004, Alabi2020}). The data for this galaxy are taken from the works of \citet{Yagi2016, Alabi2018} and \citet{FerreMateu2018}. 

\subsection{Yagi 275}
Yagi 275 is a Coma cluster galaxy. The magnitude was calculated from $R$-band using $V = R + 0.5$ (based on Virgo dEs and Coma LSBs; \citealp{vanZee2004, Alabi2020}). The data for this galaxy are taken from the works of \citet{Yagi2016, Alabi2018, Chilingarian2019} and \citet{FerreMateu2018}. 

\subsection{Yagi 276}
Yagi 276 is a Coma cluster galaxy. The magnitude was calculated from $R$-band using $V = R + 0.5$ (based on Virgo dEs and Coma LSBs; \citealp{vanZee2004, Alabi2020}). The data for this galaxy are taken from the works of \citet{Yagi2016, Alabi2018} and \citet{FerreMateu2018}. 

\subsection{Yagi 358}
Yagi 358 is a Coma cluster galaxy. The stellar mass was calculated from the absolute magnitude assuming $M_{\star} / L_{V} =2$. The data for this galaxy are taken from the works of \citet{vanDokkum2017}, \citet{Lim2018} and \citet{Gannon2023}.  

\subsection{Yagi 418}
Yagi 418 is a Coma cluster galaxy. The $M_V$ was calculated from $R$-band using $V = R + 0.5$ (based on Virgo dEs and Coma LSBs; \citealp{vanZee2004, Alabi2020}). Stellar population properties for this galaxy are presented in \cite{RuizLara2018} but here we prefer the \cite{FerreMateu2023} age/metallicity values due to their being mass-weighted in contrast to the \citet{RuizLara2018} light-weighted values. We note that the ages are in good agreement between the two studies, as is expected for such intermediate-to-old stellar populations. The data for this galaxy are taken from the works of \citet{Yagi2016, Alabi2018} and \citet{FerreMateu2018}. 

\subsection{Notable galaxies excluded from this catalogue}
Here we discuss several notable galaxies and studies that we exclude from this catalogue:

\begin{itemize}
    \item While we include 2 galaxies from the study of \citet{Chilingarian2019} that meet our UDG definition the remaining 6 are too bright and/or small to meet our UDG criteria. As such, they are excluded from this sample. 
    \item We exclude the galaxy PUDG-R24 from the study of \citet{Gannon2022} as it is too bright in surface brightness ($\langle \mu_{V} \rangle_{\rm e} \approx 24.35$~mag arcsec$^{-2}$) to meet our definition. In \citet{Gannon2022} the galaxy was considered a UDG as it was expected to fade into the UDG regime in the next few Gyr. 
    \item We exclude the galaxies OSG1 and OSG2 from \citet{RuizLara2018} due to their being light-weighted stellar population properties, rather than the mass-weighted properties presented herein. 
    \item We exclude the stacked UDG stellar population properties from \citet{Rong2020} as it is both 1) not the results for a single galaxy and 2) includes in the stack many objects that are too bright to meet our UDG definition. It is worth noting that many of these objects do have similar stellar surface densities to the UDGs in our catalogue, it is their predominantly younger stellar populations that result in their being too bright for the surface brightness criterion \citep{Rong2020}.
    \item We exclude the two galaxies presented in \citet{Greco2018} as: 1) the metallicities are lower limits and have not been measured and 2) the ages are not mean stellar ages but instead the age since the onset of star formation. We additionally note that the galaxy LSBG-285 presented by \citet{Greco2018} is too small to meet our UDG definition.
    \item  We exclude the UDGs presented in \citet{Trujillo2017} and \citet{Bellazzini2017} as only gas-phase metallicities and not stellar metallicities, are reported. We additionally note that both \citet{Bellazzini2017} galaxies are too bright to meet our UDG definition. 
    \item We exclude the galaxy NGC~1052-DF4 \citep{vanDokkum2019} from our catalogue as it does not meet the surface brightness cut of our UDG definition. To be specific, using the surface brightness at the effective radius and S\'ersic index for NGC~1052-DF4 reported in \citet[25.1 mag arcsec$^{-2}$ and 0.79 respectively]{Cohen2018} and equation 9 of \citet{Graham2005} we calculate an average surface brightness within the half-light radius of $\langle \mu_{V} \rangle_{\rm e} \approx 24.5$~mag arcsec$^{-2}$ which does not meet our definition. 
\end{itemize}

It is also worth noting that many UDGs have measurements such as redshift and rotation available from their associated HI disk (e.g., \citealt{Leisman2017, Spekkens2018, ManceraPina2019, ManceraPina2020, ManceraPina2022, Karunakaran2020, Gault2021, Kong2022, Obeirne2024}). Our chosen criteria for this catalogue do not include these galaxies as we wish to focus on the galaxies' stellar population properties, and not that of their HI. We do note that much may be learned by comparing the two properties (e.g., \citealp{KadoFong2022a, KadoFong2022b}) but that is beyond the scope of this work.

\section{Discussion} \label{sec:discussion}
\subsection{Catalogue Properties}


In Figure \ref{fig:histograms} we present histograms of catalogue parameters. Where available, we include results from the SED fitting of field and group UDGs in the MATLAS survey from \citet{Buzzo2024}. We picked this catalogue for comparison as it contains a greater number of UDGs (59) than our current work and as it has been used to argue for distinct formation pathways for UDGs through a K-means analysis. It is worth noting that the MATLAS survey primarily samples less dense field and group environments while the spectroscopic catalogue is heavily biased toward cluster environments. Moreover, the spectroscopic UDGs tend to be intrinsically brighter, have higher stellar masses, are larger, more GC-rich, older and have a wider spread in their metallicities. Spectroscopic UDGs being larger and brighter than those UDGs studied with SED fitting is likely a selection effect as it is a requirement of UDG spectroscopy for the target to be relatively bright to get meaningful results. Similar conclusions have also been drawn by \citet{Gannon2023}. 


Notably, non-UDGs that are more luminous and/or larger half-light radius galaxies tend to host richer GC systems (see e.g., \citealp{harris2017}). On average the catalogue UDG sample presented here hosts more GCs than the SED sample of Buzzo et al as may be expected as they are also on average larger and brighter. Thus it is more likely that these UDGs have formed via the ``failed galaxy" pathway that has been proposed by various authors (e.g., \citealp{Peng2016, Lim2018, Danieli2022, Forbes2024}).

\begin{figure}
    \centering
    \includegraphics[width = 0.45 \textwidth]{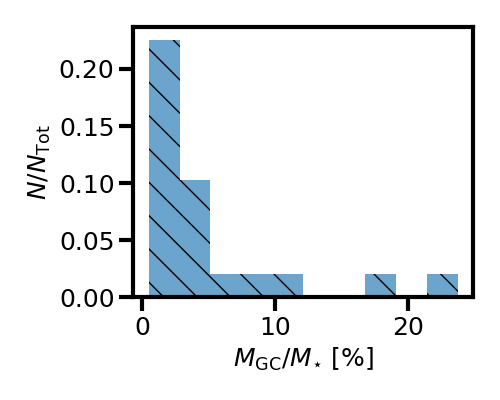}
    \caption{The percentage of stellar mass in the GC system for UDGs in the catalogue. We calculate this property from the UDGs' GC counts using a mean GC-mass of 2$\times10^{5}$~M$_\odot$. Many of the spectroscopically studied UDGs have a significant percentage of their stellar mass contained within their GC system making it likely they are `failed galaxy' UDGs.}
    \label{fig:gc_percentage}
\end{figure}

In Figure \ref{fig:gc_percentage} we plot a histogram of the percentage of stellar mass in the GC system for UDGs in the catalogue. We calculate this percentage assuming a mean GC-mass of 2$\times10^{5}$~M$_\odot$ from the stellar mass (M$_{\star}$) and GC richness ($N_{\rm GC}$) of the UDGs as:

\begin{equation}
    M_{\rm GC} / M_{\star} = \frac{2\times10^5 \times N_{\rm GC}}{M_\star}\times 100
\end{equation}

Note that for the UDGs NGC~1052-DF2 and DGSAT-I the approximation of a mean GC mass of 2$\times10^{5}$~M$_\odot$ is likely too low given the overluminous star clusters known to be associated with these galaxies. The value included in the histogram will still provide a lower limit to the percentage of their stellar mass contained within their GC system. 

In comparison to a more normal dwarf galaxy of UDG stellar mass, which has a $M_{\rm GC} / M_{\star}\approx0.5\%$~\citep{Forbes2020a},~many spectroscopically studied UDGs have extremely rich GC systems, with $>5\%$~ of their stellar mass in their GC system. In our catalogue, these galaxies are: 1) NGVSUDG-19 (5.4\%), 2) VCC~615 (8.3\%), 3) NGC~5846\_UDG1 (9.8\%), 4) VCC~615 (8.3\%) and 5) VLSB-B (23.7\%).


There is an expectation that GCs will experience significant mass loss via tidal shocking, evaporation of stars bound to the GCs and the complete dissolution of the lowest mass GCs. It is commonly thought that GC systems may lose a significant fraction ($>75\%$) of their stellar mass after initial formation \citep{Larsen2012, ReinaCampos2018}. Accounting for these processes, many UDGs with $M_{\rm GC} / M_{\star}>5\%$ are consistent with having experienced little subsequent star formation post-GC formation \citep{Danieli2022}. Due to the lack of star formation after the GC formation epoch, these may be interpreted as `failed galaxy' UDGs, possibly consistent with being pure stellar halos (e.g., \citealp{Peng2016}). 

\subsection{Catalogue Correlations}

\begin{figure*}
    \centering
    \includegraphics[width = 0.96 \textwidth]{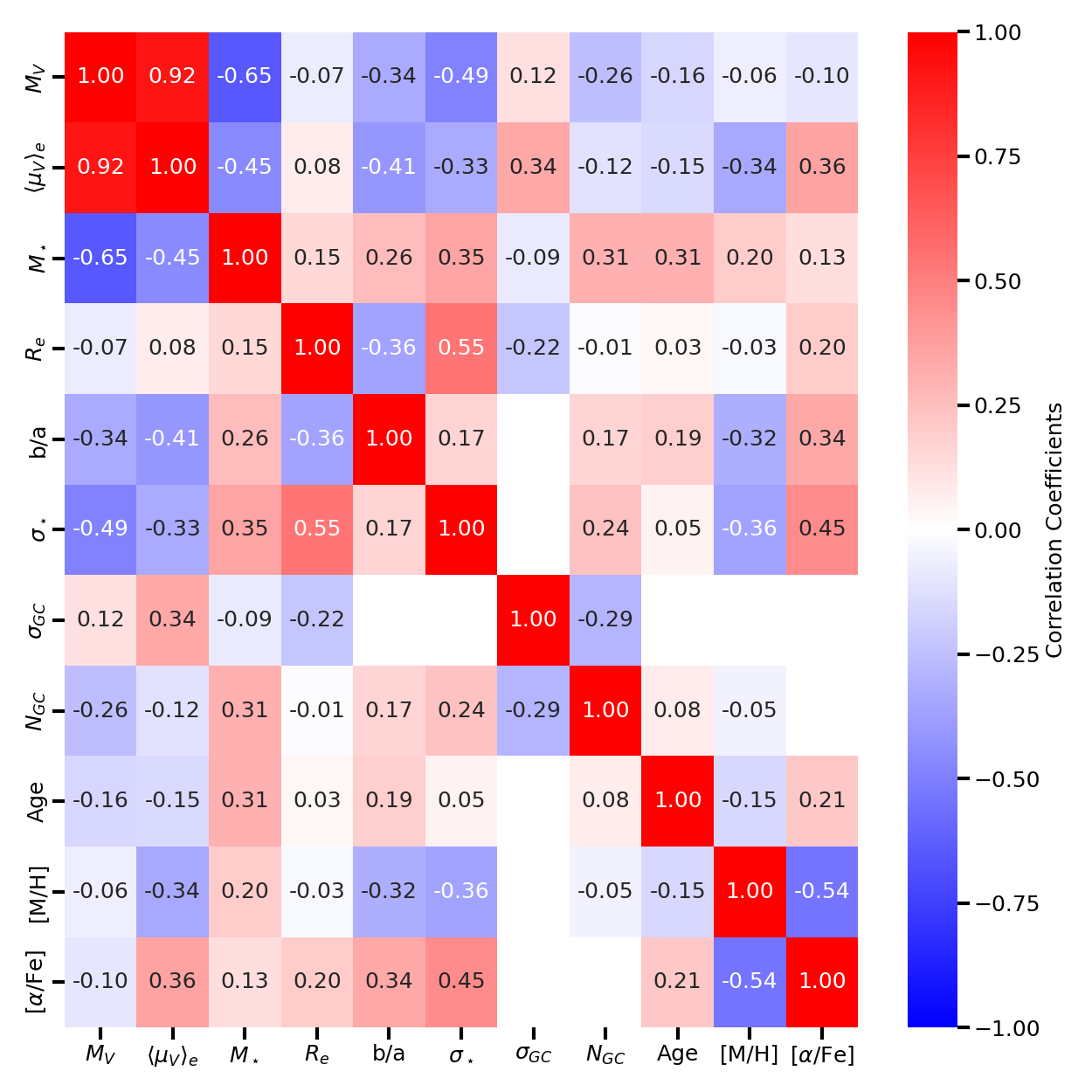}
    \caption{A heatmap of the correlation matrix for the major properties in the catalogue. Correlations values are missing when they would rely on fewer than 10 datapoints for calculation. The majority of our properties are not correlated. For a full discussion of the interesting correlations found in the correlation matrix (i.e., those with $|$correlation coefficient$|>0.5$), please refer to the text.}
    \label{fig:correlation}
\end{figure*}

\begin{figure*}
    \centering
    \includegraphics[width = 0.98 \textwidth]{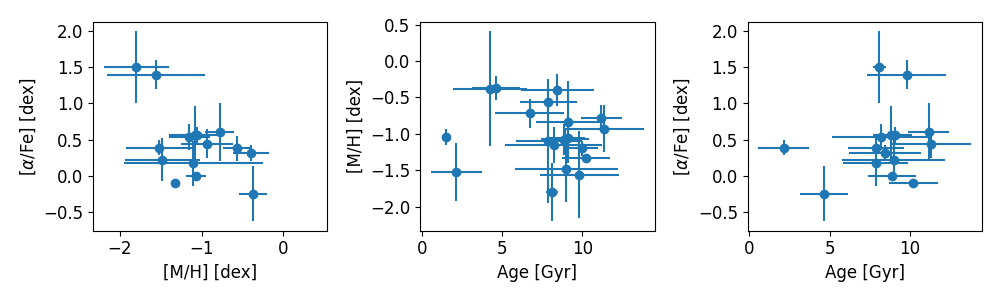}
    \caption{\textit{Left:} [$\alpha$/Fe] \textit{vs.} [M/H] for the catalogued galaxies. A correlation is found between these two parameters. \textit{Centre:} [M/H] \textit{vs.} mean stellar age for the catalogued galaxies. No correlation is found between these two parameters. \textit{Right:} [$\alpha$/Fe] \textit{vs.} mean stellar age for the catalogued galaxies. No correlation is found between these two parameters. The lack of an age -- metallicity correlation is likely due to the presence of two outliers at low ages and metallicities that do not follow a standard age--metallicity relationship.}
    \label{fig:3corr}
\end{figure*}

In Figure \ref{fig:correlation} we show the correlation matrix of the major properties included in the catalogue. We require each correlation to have 10 entries in the intersection of their parameters to calculate its coefficient. The vast majority of the properties are not correlated with coefficients between -0.5 and 0.5. Four correlations with $|$correlation coefficient$|>0.5$ are found. We have checked and all these correlations remain if we exclude the two much fainter galaxies in the sample i.e.,  Andromeda XIX and Antlia II, for which analogues are likely not readily observable beyond the Local Group. The correlations found are:
\begin{enumerate}
    \item Between $M_{V}$ and $\langle \mu_V \rangle_e$. UDGs with higher luminosities also tend to exhibit higher fluxes. This is as expected.
    \item Between the stellar mass ($M_\star$) and $M_V$. Here the correlation coefficient is negative due to the nature of the magnitude system. UDGs that are more luminous also tend to exhibit higher stellar masses. This is as expected.
    \item Between the stellar sigma ($\sigma_\star$) and the half-light radius ($R_{\rm e}$). UDGs that have higher stellar sigma are dynamically hotter and tend to be larger. This is expected given the fundamental plane of elliptical galaxies and provides support for predicting UDG velocity dispersions via the fundamental plane (e.g., \citealp{Zaritsky2023a, Zaritsky2023b}).
    \item Between the alpha element abundance ([$\alpha$/Fe]) and the metallicity ([M/H]). UDGs that are more alpha-enhanced also tend to be lower in overall metallicity. A similar trend was found by \citet{FerreMateu2023} from which much of our data is sourced. The leading line of reasoning to explain this trend is that observed UDGs cover a small stellar mass range. Thus, those that formed this stellar mass quickly in the early Universe will have elevated alpha abundances and low metallicities reflective of this early, fast formation. They will not experience significant subsequent star formation to change these metallicities as any significant subsequent star formation would cause them to not fulfil the UDG definition.
    
    Under this line of reasoning, there is likely an expectation that there will also be a correlation between age and either alpha abundance/metallicity, which is not found in our catalogue. We show the [$\alpha$/Fe] -- [M/H] correlation, along with the [M/H] -- mean stellar age and [$\alpha$/Fe] -- mean stellar age non-correlations in Figure \ref{fig:3corr}. When looking at the centre panel, it is possible that a correlation is not found between age and metallicity due to the two galaxies at low age and metallicity. If these galaxies were removed the remaining galaxies would follow a standard age -- metallicity relationship. Alternatively, the lack of trends may suggest the need for new formation pathways to be considered. e.g., the UDG DGSAT-I has both an elevated alpha abundance and signs of recent star formation \citep{MartinNavarro2019, Janssens2022} which does not fit our line of reasoning for a `failed galaxy' UDG.
\end{enumerate}


\subsection{Catalogue UDG Populations}
Finally, it was possible to split the \citet{Buzzo2024} UDG sample using the machine learning K-means method into two samples that resembled the expected properties for `failed galaxy' UDGs and `puffy dwarf' UDGs. We have attempted to perform a K-means analysis on the UDGs presented in this work to similarly split them into `puffy dwarfs' and `failed galaxies' but found that it was not applicable. We base this on measuring the silhouette score of the calculated K-means clusters as a function of the number of clusters found. The silhouette score is a measure of how similar an object is to its assigned cluster with values ranging from -1 to 1. In general, silhouette scores $>0.7$ are required for a clustering to be considered `strong'. When splitting into 2 clusters (i.e., the expectation of a `puffy dwarf'/`failed galaxy' dichotomy) the clustering is at best very weak (i.e., silhouette score $<0.3$). The addition of more K-means clusters does not solve this issue. We conclude that it is currently not warranted to segment the current spectroscopic data presented herein into separate, distinct UDG populations. We suggest this should be kept in mind when extrapolating the findings of current spectroscopic UDG studies more generally to the entire population.

\section{Catalogue Access and Citing} \label{sec:housekeeping}

\begin{figure}
    \centering
    \includegraphics[width = 0.3 \textwidth]{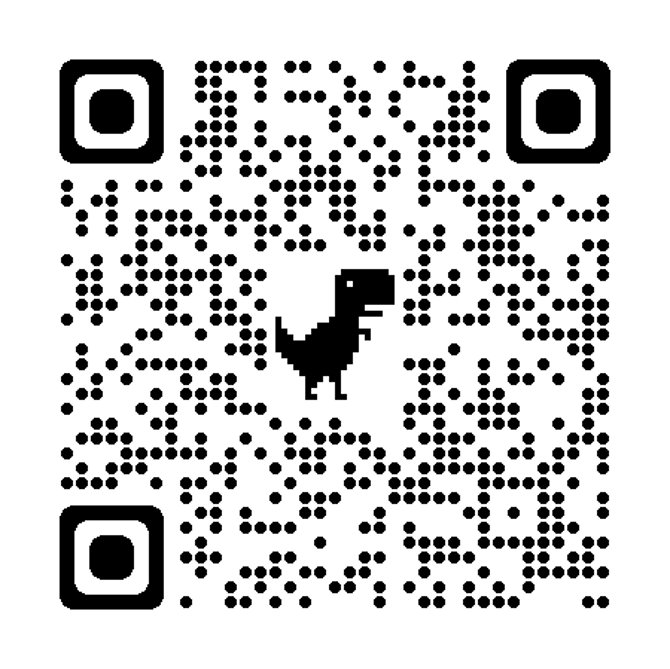}
    \caption{A QR code that you may scan to take you to the online catalogue.}
    \label{fig:QR}
\end{figure}

The catalogue described above has been made publicly available via the \texttt{GitHub} of the first author \here. We include a QR code that will take the reader of this work to the catalogue in Figure \ref{fig:QR}. As part of the online catalogue a \texttt{.bib} \texttt{LaTeX} file is included which holds citations of all works that have contributed to this catalogue. It has been requested by community members via discussions at \textit{The Sunrise of Ultra-Diffuse Galaxies} conference in Sesto, Italy, July 2023 that individual works contributing to this catalogue are cited when it is used. To facilitate this request a \texttt{LaTeX} input that should work with the provided \texttt{.bib} \texttt{LaTeX} file and the \texttt{natbib} package are included in the online catalogue. For reference, we include it below:

\texttt{\\ $\backslash$cite\{mcconnachie2012, vanDokkum2015, Beasley2016, Martin2016, Yagi2016, MartinezDelgado2016, vanDokkum2016, vanDokkum2017, Karachentsev2017, vanDokkum2018 Toloba2018, Gu2018, Lim2018, RuizLara2018, Alabi2018, FerreMateu2018, Forbes2018, MartinNavarro2019, Chilingarian2019, Fensch2019, Danieli2019, vanDokkum2019b, torrealba2019, Iodice2020, Collins2020, Muller2020, Gannon2020, Lim2020, Muller2021, Forbes2021, Shen2021, Gannon2021, Gannon2022, Mihos2022, Danieli2022, Villaume2022, Webb2022, Saifollahi2022, Janssens2022, Gannon2023, FerreMateu2023, Toloba2023, Iodice2023, Shen2023\}}\\

We intend to continue to update the online version of the catalogue and reference list described herein as new UDG works are released. It is therefore advisable to include a date of retrieval when using these data. If we have missed data please contact the author for correspondence JSG (jonah.gannon@gmail.com) so that we may include it in this catalogue. 

\section{Conclusions} \label{sec:conclusions}
In this work, we have presented a literature compilation of UDG spectroscopic data along with the details to access it online. In comparison to the SED fitting of a larger UDG sample from the MATLAS survey we find the galaxies in our catalogue tend to be intrinsically brighter, have higher stellar mass, are larger, more GC-rich, older and have a wider spread in their metallicities. Spectroscopically studied UDGs also tend to be in denser, cluster environments while the SED sample is biased to groups and the field. These biases should be kept in mind when using UDG spectroscopic data to draw broad conclusions on the formation of the populations as a whole. 

We show that many UDGs in this catalogue have a significant fraction of their stellar mass bound within their GC system. In current models for GC evolution, this may leave little room for star formation after the initial cluster formation epoch as much of their non-GC stellar mass can be explained as the product of GC dissolution/evaporation.

We investigate the correlations of major properties within the catalogue, finding the majority are uncorrelated. Of most interest is that alpha abundance and total metallicity are anti-correlated. UDGs that are more alpha-enhanced tend to have lower metallicity. This may be expected if some UDGs form fast and early when the Universe is less metal-enriched. Under this expectation, similar trends with age may be expected, but these are not found. We are currently unable to comment on whether this is related to the underlying formation pathways of UDGs or simply a result of outliers and low number statistics in the data.

Finally, we note that we are unable to reproduce the machine learning, K-means results of UDGs with SED fitting. The UDGs in our catalogue do not cluster strongly in K-space and do not cluster as distinctly as those studied in SED fitting. It is currently not warranted to separate the spectroscopically studied UDGs into multiple sub-populations.

Those wishing to use our catalogue may access it \here~or by scanning the QR code in Figure \ref{fig:QR}. We intend to keep this catalogue updated beyond the publication of this paper.

\section*{Acknowledgements}
We thank the anonymous referee for their swift, helpful review of our manuscript. We thank J. Pfeffer, W. Couch, S. Janssens, E. Peng and I. Trujillo for insightful conversations that helped motivate this work. We thank those who helped us observe much of the data that is included in our contribution to the catalogue. JSG thanks the Instituto de Astrofisica de Canarias for their early career visitor program which supported him while working on part of this work. This research was partially supported by the Instituto de Astrofisica de Canarias through their Early Career Visitor Program. AFM acknowledges support from RYC2021-031099-I and PID2021-123313NA-I00 of MICIN/AEI/10.13039/501100011033/FEDER,UE,NextGenerationEU/PRT. AJR was supported by National Science Foundation grant AST-2308390. This work was partially supported by a NASA Keck PI Data Award, administered by the NASA Exoplanet Science Institute. This research was supported by the Australian Research Council Centre of Excellence for All Sky Astrophysics in 3 Dimensions (ASTRO 3D), through project number CE170100013. DF and JB thank ARC DP220101863 for financial support.

\section*{Data Availability}
This paper comprises a publicly released collection of literature data. The catalogue is available at \url{https://github.com/gannonjs/Published_Data/tree/main/UDG_Spectroscopic_Data}.



\bibliographystyle{mnras}
\bibliography{bibliography} 




\appendix

\section{Catalogue Tables}
Here we provide tables that display the catalogue at the time of paper publication. 

\begin{table*}
    \centering
    \begin{tabular}{lllllllll}
    \hline
    Name & Other Names & Environment & Distance & $M_{V}$ & $\langle \mu_{V} \rangle_{\rm e, circ}$ &$M_{\star}$ & $R_{\rm e}$ & $b/a$\\
    & & & [Mpc] & [mag] & [mag arcsec$^{-2}$] & [$\times10^{8}$~M$_{\odot}$] & [kpc] \\
    \hline
    Andromeda XIX & LEDA 5056919 & 2 & 0.93 & $-$10 & 31.0 & 0.016 & 3.1 & 0.42 \\
    Antlia II &  & 2 & 0.132 & $-$9.03 & 31.9 & 0.0088 & 2.9 & 0.62\\
    DF44 & Dragonfly 44, Yagi011 & 1 & 100 & $-$16.2 & 25.7 & 3 & 4.7 & 0.69 \\
    DF07 & Yagi680 & 1 & 100 & $-$16.2 & 25.6 & 4.35 & 4.3 & 0.76 \\
    DF17 & Yagi165 & 1 & 100 & $-$15.3 & 25.49 & 2.63 & 4.4 & 0.71 \\
    DF26 & Yagi093; GMP2748 & 1 & 100 & $-$15.64 & 25.22 & 3.05 & 3.5 & 0.68 \\
    DFX1 & Yagi013; GMP2175 & 1 & 100 & $-$15.8 & 25.5 & 3.4 & 3.5 & 0.62 \\
    DGSAT-$\mathrm{I}$ &  & 3 & 78 & $-$16.3 & 25.6 & 4 & 4.7 & 0.87 \\
    Hydra $\mathrm{I}$ UDG 11 &  & 1 & 51 & $-$14.62 & 25.04 & 0.63 & 1.66 & 0.92 \\
    J130026.26+272735.2 & GMP 2673 & 1 & 100 & $-$16.27 & 24.83 & 1.57 & 3.7 & $-$999 \\
    NGC 1052-DF2 & \makecell{RCP 29; [KKS2000] 04;\\ LEDA 3097693; Ta21-12200} & 2 & 21.7 & $-$15.3 & 24.8 & 2 & 2.2 & 0.85 \\
    NGC 5846\_UDG1 & MATLAS-2019; NGC 5846-156 & 2 & 26.5 & $-$15 & 25.2 & 1.1 & 2.14 & 0.9 \\
    NGVSUDG-19 &  & 1 & 16.5 & $-$13.8 & 26.37 & 0.62 & 2.18 & $-$999 \\
    NGVSUDG-20 &  & 1 & 16.5 & $-$13.2 & 27.94 & 0.13 & 3.48 & $-$999 \\
    PUDG-R15 &  & 1 & 75 & $-$15.65 & 24.83 & 2.59 & 2.5 & 0.97 \\
    PUDG-R16 &  & 1 & 75 & $-$15.9 & 25.4 & 5.75 & 4.2 & 0.7 \\
    PUDG-R84 &  & 1 & 75 & $-$15.4 & 24.68 & 2.2 & 2.0 & 0.97 \\
    PUDG-S74 &  & 1 & 75 & $-$16.49 & 24.82 & 7.85 & 3.8 & 0.86 \\
    Sagittarius dSph &  & 2 & 0.02 & $-$15.5 & 25.13 & 1.32 & 2.6 & 0.48 \\
    UDG1137+16 & dw1137+16 & 2 & 21 & $-$14.65 & 26.55 & 1.4 & 3.3 & 0.8 \\
    VCC 1017 & NGVSUDG-09; LEDA40869 & 1 & 16.5 & $-$16.7 & 24.89 & 3.35 & 4.29 & $-$999\\
    VCC 1052 & NGVSUDG-10; LEDA40932 & 1 & 16.5 & $-$15.2 & 26.13 & 2.08 & 3.79 & $-$999\\
    VCC 1287 & NGVSUDG-14; LEDA41311 & 1 & 16.5 & $-$15.6 & 25.71 & 2 & 3.7 & 0.8 \\
    VCC 615 & NGVSUDG-A04; LEDA40181 & 1 & 17.7 & $-$14.2 & 26 & 0.73 & 2.3 & $-$999\\
    VCC 811 & NGVSUDG-05; LEDA40541 & 1 & 16.5 & $-$14.3 & 26.3 & 0.73 & 2.71 & $-$999\\
    VLSB-B & NGVSUDG-11 & 1 & 12.7 & $-$12.3 & 27.6 & 0.22 & 1.9 & 0.83 \\
    VLSB-D & NGVSUDG-04 & 1 & 16.5 & $-$13.7 & 26.85 & 0.58 & 13.4 & 0.45 \\
    WLM &  & 2 & 0.93 & $-$14.25 & 26.16 & 0.41 & 2.11 & 0.35\\
    Yagi098 &  & 1 & 100 & $-$14.6 & 25.64 & 1.07 & 2.3 & 0.88 \\
    Yagi275 & GMP3418; J125929.89+274303.0 & 1 & 100 & $-$15.3 & 24.83 & 0.94 & 2.9 & 0.49 \\
    Yagi276 & DF28 & 1 & 100 & $-$14.86 & 25.37 & 1.41 & 2.25 & 0.91 \\
    Yagi358 & Y358; GMP3651 & 1 & 100 & $-$14.8 & 25.6 & 1.38 & 2.3 & 0.83 \\
    Yagi418 &  & 1 & 100 & $-$14.11 & 25.19 & 1.24 & 1.58 & 0.79\\
     \hline
    \end{tabular}%
    \caption{The first 8 columns of the full online catalogue. From left to right these are: 1) Primary Name, 2) Other names,  3) Environment where 1 = Cluster, 2 = Group and 3 = Field, 4) Distance noting that this is frequently assumed based on environmental association, 5) $V$-band absolute magnitude, 6) the average $V$-band surface brightness within the half-light radius, 7) Stellar mass, 8) Semi-major half-light radius and 9) Axial ratio, $b/a$. When values are not available they are listed as $-$999. The full table is available online \here.}
    \label{tab:cropped_table1}
\end{table*}

\begin{table*}
    \centering
    \resizebox{\textwidth}{!}{%
    \begin{tabular}{lllllllllllll}
    \hline
    Name & $V_r$ & $V_r+$ & $V_r-$ & $\sigma_\star$ & $\sigma_\star+$ & $\sigma_\star-$ & $\sigma_{\rm GC}$ & $\sigma_{\rm GC}+$ & $\sigma_{\rm GC}-$ & $N_{\rm GC}$ & $N_{\rm GC}+$ & $N_{\rm GC}-$ \\
    & [km s$^{-1}$] & [km s$^{-1}$] & [km s$^{-1}$] & [km s$^{-1}$] & [km s$^{-1}$] & [km s$^{-1}$] & [km s$^{-1}$] & [km s$^{-1}$] & [km s$^{-1}$] & & & \\
    \hline
    Andromeda XIX & $-$109 & 1.6 & 1.6 & 7.8 & 1.7 & 1.5 & $-$999 & $-$999 & $-$999 & $-$999 & $-$999 & $-$999 \\
    Antlia II & 290.7 & 1.5 & 1.5 & 5.71 & 1.08 & 1.08 & $-$999 & $-$999 & $-$999 & $-$999 & $-$999 & $-$999 \\
    DF44 & 6234 & $-$999 & $-$999 & 33 & 3 & 3 & $-$999 & $-$999 & $-$999 & 74 & 18 & 18 \\
    DF07 & 6600 & 40 & 26 & $-$999 & $-$999 & $-$999 & $-$999 & $-$999 & $-$999 & 23 & 7 & 7 \\
    DF17 & 8315 & 43 & 43 & $-$999 & $-$999 & $-$999 & $-$999 & $-$999 & $-$999 & 27 & 4 & 4 \\
    DF26 & 6611 & 137 & 137 & $-$999 & $-$999 & $-$999 & $-$999 & $-$999 & $-$999 & 20 & 20.7 & 20.7 \\
    DFX1 & 8107 & $-$999 & $-$999 & 30 & 7 & 7 & $-$999 & $-$999 & $-$999 & 62 & 17 & 17 \\
    DGSAT-$\mathrm{I}$ & 5439 & 8 & 8 & 56 & 10 & 10 & $-$999 & $-$999 & $-$999 & 12 & 2 & 2 \\
    Hydra $\mathrm{I}$ UDG 11 & 3507 & 3 & 3 & 20 & 8 & 8 & $-$999 & $-$999 & $-$999 & 7 & 3 & 3 \\
    J130026.26+272735.2 & 6939 & 2 & 2 & 19 & 5 & 5 & $-$999 & $-$999 & $-$999 & $-$999 & $-$999 & $-$999 \\
    NGC 1052-DF2 & 1805 & 1.1 & 1.1 & 8.5 & 2.3 & 3.1 & 7.8 & 5.2 & 2.2 & 7.1 & 7.33 & 4.34 \\
    NGC 5846\_UDG1 & 2167 & 2 & 2 & 17 & 2 & 2 & 9.4 & 7 & 5.4 & 54 & 9 & 9 \\
    NGVSUDG-19 & 296 & 37 & 38 & $-$999 & $-$999 & $-$999 & 61 & 47 & 23 & 16.8 & 7.5 & 7.5 \\
    NGVSUDG-20 & 946 & 42 & 41 & $-$999 & $-$999 & $-$999 & 89 & 42 & 27 & 11.3 & 8.6 & 8.6 \\
    PUDG-R15 & 4762 & 2 & 2 & 10 & 4 & 4 & $-$999 & $-$999 & $-$999 & $-$999 & $-$999 & $-$999 \\
    PUDG-R16 & 4679 & 2 & 2 & 12 & 3 & 3 & $-$999 & $-$999 & $-$999 & $-$999 & $-$999 & $-$999 \\
    PUDG-R84 & 4039 & 2 & 2 & 19 & 3 & 3 & $-$999 & $-$999 & $-$999 & $-$999 & $-$999 & $-$999 \\
    PUDG-S74 & 6215 & 2 & 2 & 22 & 2 & 2 & $-$999 & $-$999 & $-$999 & $-$999 & $-$999 & $-$999 \\
    Sagittarius dSph & 140 & 2 & 2 & 11.4 & 0.7 & 0.7 & $-$999 & $-$999 & $-$999 & 8 & 0 & 0 \\
    UDG1137+16 & 1014 & 3 & 3 & 15 & 4 & 4 & $-$999 & $-$999 & $-$999 & $-$999 & $-$999 & $-$999 \\
    VCC 1017 & 38 & 31 & 33 & $-$999 & $-$999 & $-$999 & 83 & 33 & 22 & 16.5 & 11.2 & 11.2 \\
    VCC 1052 & $-$292 & 6 & 7 & $-$999 & $-$999 & $-$999 & 6 & 11 & 4 & 17.9 & 11.5 & 11.5 \\
    VCC 1287 & 1116 & 2 & 2 & 19 & 6 & 6 & 35 & 12 & 12 & 22 & 8 & 8 \\
    VCC 615 & 2089 & 16 & 2.7 & $-$999 & $-$999 & $-$999 & 36 & 22 & 18 & 30.3 & 9.6 & 9.6 \\
    VCC 811 & 982 & 29 & 29 & $-$999 & $-$999 & $-$999 & 64 & 33 & 19 & 15.8 & 8.4 & 8.4 \\
    VLSB-B & 40 & 14 & 14 & $-$999 & $-$999 & $-$999 & 45 & 14 & 10 & 26.1 & 9.9 & 9.9 \\
    VLSB-D & 1035 & 6 & 5 & $-$999 & $-$999 & $-$999 & 12 & 6 & 6 & 13 & 6.9 & 6.9 \\
    WLM & $-$130 & 1 & 1 & 17.5 & 2 & 2 & $-$999 & $-$999 & $-$999 & 1 & 0 & 0 \\
    Yagi098 & 5980 & 82 & 82 & $-$999 & $-$999 & $-$999 & $-$999 & $-$999 & $-$999 & $-$999 & $-$999 & $-$999 \\
    Yagi275 & 4847 & 4 & 4 & 23 & 6 & 6 & $-$999 & $-$999 & $-$999 & $-$999 & $-$999 & $-$999 \\
    Yagi276 & 7343 & 102 & 102 & $-$999 & $-$999 & $-$999 & $-$999 & $-$999 & $-$999 & $-$999 & $-$999 & $-$999 \\
    Yagi358 & 7969 & 2 & 2 & 19 & 3 & 3 & $-$999 & $-$999 & $-$999 & 28 & 5.3 & 5.3 \\
    Yagi418 & 8335 & 187 & 187 & $-$999 & $-$999 & $-$999 & $-$999 & $-$999 & $-$999 & $-$999 & $-$999 & $-$999 \\
    \hline
    \end{tabular}
    }
    \caption{The subsequent 12 columns of our online catalogue. From left to right these are: 1) Primary Name, 2) Recessional velocity ($V_{r}$), 3) the positive uncertainty in the recessional velocity, 4) the negative uncertainty in the recessional velocity, 5) the stellar velocity dispersion ($\sigma_{\star}$) 6) the positive uncertainty in the stellar velocity dispersion, 7) the negative uncertainty in the stellar velocity dispersion, 8) the measured velocity dispersion of the GC system ($\sigma_{\rm GC}$), 9) the positive uncertainty in the GC velocity dispersion, 10) the negative uncertainty in the GC velocity dispersion, 11) the total number of associated GCs, 12) the positive uncertainty in the total GC number, 13) the negative uncertainty in the total GC number. When values are not available they are listed as $-$999. The full table is available online \here. }
    \label{tab:cropped_table2}
\end{table*}

\begin{table*}
    \centering
    \begin{tabular}{llllllllll}
    \hline
    Name & Age & Age+ & Age$-$ & [M/H] & [M/H]+ & [M/H]$-$ & [$\alpha$/Fe] & [$\alpha$/Fe]+ & [$\alpha$/Fe]$-$ \\
    & [Gyr] & [Gyr] & [Gyr] & [dex] & [dex] & [dex] & [dex] & [dex] & [dex] \\
    \hline
    Andromeda XIX & $-$999 & $-$999 & $-$999 & $-$999 & $-$999 & $-$999 & $-$999 & $-$999 & $-$999 \\
    Antlia II & $-$999 & $-$999 & $-$999 & $-$999 & $-$999 & $-$999 & $-$999 & $-$999 & $-$999 \\
    DF44 & 10.23 & 1.5 & 1.5 & $-$1.33 & 0.05 & 0.04 & $-$0.10 & 0.06 & 0.06 \\
    DF07 & 11.18 & 1.27 & 1.27 & $-$0.78 & 0.18 & 0.18 & 0.6 & 0.4 & 0.4 \\
    DF17 & 9.11 & 2 & 2 & $-$0.83 & 0.56 & 0.51 & $-$999 & $-$999 & $-$999 \\
    DF26 & 7.88 & 1.76 & 1.76 & $-$0.56 & 0.18 & 0.18 & 0.38 & 0.17 & 0.17 \\
    DFX1 & 8.84 & 1.13 & 1.13 & $-$1.08 & 0.21 & 0.21 & 0.57 & 0.4 & 0.4 \\
    DGSAT-$\mathrm{I}$ & 8.1 & 0.4 & 0.4 & $-$1.8 & 0.4 & 0.4 & 1.5 & 0.5 & 0.5 \\
    Hydra $\mathrm{I}$ UDG 11 & 10 & 1 & 1 & $-$1.2 & 0.1 & 0.1 & $-$999 & $-$999 & $-$999 \\
    J130026.26+272735.2 & 1.5 & 0.1 & 0.1 & $-$1.04 & 0.11 & 0.11 & $-$999 & $-$999 & $-$999 \\
    NGC 1052-DF2 & 8.9 & 1.5 & 1.5 & $-$1.07 & 0.12 & 0.12 & 0 & 0.05 & 0.05 \\
    NGC 5846\_UDG1 & 8.2 & 3.05 & 3.05 & $-$1.15 & 0.25 & 0.25 & 0.54 & 0.18 & 0.18 \\
    NGVSUDG-19 & $-$999 & $-$999 & $-$999 & $-$999 & $-$999 & $-$999 & $-$999 & $-$999 & $-$999 \\
    NGVSUDG-20 & $-$999 & $-$999 & $-$999 & $-$999 & $-$999 & $-$999 & $-$999 & $-$999 & $-$999 \\
    PUDG-R15 & 11.32 & 2.52 & 2.52 & $-$0.93 & 0.32 & 0.32 & 0.44 & 0.2 & 0.2 \\
    PUDG-R16 & $-$999 & $-$999 & $-$999 & $-$999 & $-$999 & $-$999 & $-$999 & $-$999 & $-$999 \\
    PUDG-R84 & 8.99 & 3.2 & 3.2 & $-$1.48 & 0.46 & 0.46 & 0.22 & 0.3 & 0.3 \\
    PUDG-S74 & 8.44 & 2.26 & 2.26 & $-$0.4 & 0.22 & 0.22 & 0.32 & 0.11 & 0.11 \\
    Sagittarius dSph & $-$999 & $-$999 & $-$999 & $-$999 & $-$999 & $-$999 & $-$999 & $-$999 & $-$999 \\
    UDG1137+16 & 2.13 & 1.58 & 1.58 & $-$1.52 & 0.4 & 0.4 & 0.39 & 0.1 & 0.1 \\
    VCC 1017 & $-$999 & $-$999 & $-$999 & $-$999 & $-$999 & $-$999 & $-$999 & $-$999 & $-$999 \\
    VCC 1052 & $-$999 & $-$999 & $-$999 & $-$999 & $-$999 & $-$999 & $-$999 & $-$999 & $-$999 \\
    VCC 1287 & 9.09 & 1.07 & 1.07 & $-$1.06 & 0.34 & 0.34 & 0.56 & 0.11 & 0.11 \\
    VCC 615 & $-$999 & $-$999 & $-$999 & $-$999 & $-$999 & $-$999 & $-$999 & $-$999 & $-$999 \\
    VCC 811 & $-$999 & $-$999 & $-$999 & $-$999 & $-$999 & $-$999 & $-$999 & $-$999 & $-$999 \\
    VLSB-B & $-$999 & $-$999 & $-$999 & $-$999 & $-$999 & $-$999 & $-$999 & $-$999 & $-$999 \\
    VLSB-D & $-$999 & $-$999 & $-$999 & $-$999 & $-$999 & $-$999 & $-$999 & $-$999 & $-$999 \\
    WLM & $-$999 & $-$999 & $-$999 & $-$999 & $-$999 & $-$999 & $-$999 & $-$999 & $-$999 \\
    Yagi098 & 6.72 & 2.16 & 2.16 & $-$0.72 & 0.2 & 0.2 & $-$999 & $-$999 & $-$999 \\
    Yagi275 & 4.63 & 1.5 & 1.5 & $-$0.37 & 0.17 & 0.17 & $-$0.25 & 0.38 & 0.38 \\
    Yagi276 & 4.24 & 2.32 & 2.32 & $-$0.38 & 0.79 & 0.79 & $-$999 & $-$999 & $-$999 \\
    Yagi358 & 9.81 & 2.46 & 2.46 & $-$1.56 & 0.6 & 0.6 & 1.4 & 0.2 & 0.2 \\
    Yagi418 & 7.87 & 2.02 & 2.02 & $-$1.1 & 0.85 & 0.85 & 0.17 & 0.31 & 0.31 \\
    \hline
    \end{tabular}%

    \caption{The subsequent 9 columns of our online catalogue. From left to right these are: 1) Primary Name, 2) Mass-weighted age, 3) Positive uncertainty in the mass-weighted age, 4) Negative uncertainty in the mass-weighted age, 5) Total mass-weighted metallicity, 6) Positive uncertainty in the total mass-weighted metallicity, 7) Negative uncertainty in the total mass-weighted metallicity, 8) Stellar alpha-abundance, 9) Positive uncertainty in the stellar alpha-abundance, 10) Negative uncertainty in the stellar alpha-abundance. When values are not available they are listed as $-$999. The full table is available online \here.  }
    \label{tab:cropped_table3}
\end{table*}


\bsp	
\label{lastpage}
\end{document}